\documentclass [a4paper,12pt,epsfig]{article}
\usepackage{latexsym}
\usepackage{graphicx}

\begin{document}

\begin{center}
{\large \bf Energy dependent sputtering of nano-clusters from a 
nanodisperse target and embedding of nanoparticles into a substrate}\\
\end{center}

\begin{center}
B. Satpati$^{1}$$^{*}$, J. Ghatak$^{1}$, B. Joseph$^{1}$, P. V. Satyam$^{1}$, 
T. Som$^{1}$, D. Kabiraj$^{2}$ and B. N. Dev$^{1}$\\

\vspace{0.2in}
$^{1}$Institute of Physics, Sachivalaya Marg, Bhubaneswar 751 005, India\\
$^{2}$Nuclear Science Centre, Aruna Asaf Ali Marg, New Delhi 110 067, India
\end{center}

\vspace{0.2in}
\begin{center}
\textbf{ABSTRACT}
\end{center}

Au nanoparticles, prepared by thermal evaporation under high vacuum
condition on Si substrate, are irradiated with Au ions at different ion
energies. During ion irradiation, embedding of nanoparticles as well as
ejection of nano-clusters is observed. Ejected particles (usually smaller
than those on the Si substrate) due to sputtering are collected on
carbon-coated transmission electron microscopy (TEM) grids. Both the TEM
grids and the ion-irradiated samples are analyzed with TEM. Unirradiated
as well as irradiated samples are also analyzed by Rutherford
backscattering spectrometry (RBS). In the case of low energy (32 keV)
ions, where the nuclear energy loss is dominant, both sputtering and
embedding are less compared to medium energy (1.5 MeV). In the high energy
regime (100 MeV), where the electronic energy loss is dominant, sputtering
is maximum but practically there is no embedding. Ion bombardment of
surfaces at an angle with respect to the surface-normal produces enhanced
embedding compared to normal-incidence bombardment. The depth of embedding
increases with larger angle of incidence. Au nanoparticles after ion
irradiation form embedded gold-silicide. Size distribution of the
sputtered Au clusters on the TEM grids for different ion energy regimes
are presented. \\

\noindent\textbf{$^{\ast }$Corresponding author} e-mail: biswarup@iopb.res.in,
Fax: 91-674-2300142.\\

\newpage
\section {Introduction}

Sputtering from solids interacting with energetic particles is an active
research area of great fundamental and applied interest. Momentum transfer
to target atoms can be accomplished through elastic collisions or
electronic processes. Linear collision cascade theory \cite{Sigmund} has
been used most successfully in describing the sputtering of metals by ions
of keV energies. Enhancements due to elastic collision spikes are observed
when both projectile and target consist of very high Z materials
\cite{Bay}.\\

\noindent When an energetic particle passes into a solid, collisions
induce several processes occur such as recoil and sputtering of
constituent atoms, defect formation, electron excitation and emission, and
photon emission. The sputtering process, especially emission process of
secondary ions, has been widely studied for various target materials under
bombardment of heavy ions. Most of studies are, however, concerned with
secondary ion mass spectrometry (SIMS) at nuclear-collision dominant low
energies [3-9]. In a MeV-energy range, an electronic-energy-loss process
becomes dominant, \cite{Ziegler} and the basic process of secondary ion
emission is different from that in the low energy range because the
electronic behavior strongly depends on the solid state property.\\

\noindent Birtcher {\it et al.} \cite{Birtcher} reported ejection of
nanoparticles from Au films that were irradiated with 50 - 400 keV Xe
ions, providing evidence for pulsed plastic flow on the material
structure. Rehn {\it et al.} \cite{Rehn} reported the size distributions
of the large clusters ($\ge$ 500 atoms) that are sputtered from the
surface of thick films by high energy ion impacts (400 and 500 keV: Ne,
Ar, Kr, and Au). They proposed that the clusters are produced when shock
waves, generated by subsurface displacement cascades, impact and ablate
the surface, as predicted in the model of Bitensky and Parilis
\cite{Bitensky}. Quite recently, a more representative set of data on
sputtering of metal targets (Au and Ag films with thickness of $\sim$1000
nm) with cluster ions of Au$_n$ (n = 1-5 \cite{Andersen}, n =1-13
\cite{Bouneau}) with the energy from 20 keV/at. to 5 MeV/at. was obtained.
Sputtering phenomenon can be expected to be considerably stronger in small
finite systems like clusters \cite{Kissel,Baranov,Satpati1}. The
bombardment of Au clusters deposited on native-oxide/Si (100) surfaces has
been investigated experimentally with ion energy and impact angle. Baranov
{\it et al.} \cite{Baranov} reported sputtering of nanodispersed targets
by 6 MeV Au$_5$ cluster ions (1.2 MeV/at.) and found that the size
distribution of desorbed clusters are systematically shifted towards
smaller sizes by comparison with the grain size distributions on the
targets. Theoretical studies by Shapiro and Tombrello \cite{Shapiro}
showed that both the ballistic and thermal spike phases of collision
cascades contribute significantly to the non-linearity and the large
sputtering yields observed in the bombardment of gold targets with small
gold clusters.\\

\noindent Ion beam mixing is a well-known technique for the formation of
metastable and/or equilibrium phases in thin film structures [20-22].  
The formation of silicides caused by the penetration of energetic ions
through the interface between a metal-film and silicon has long back been
recognized as one of the aspects of ion beam induced reactions in thin
film structures \cite{Tsaur,Poate}. Lau {\it et al.} \cite{Lau} reported
about the ion beam mixing of four metal-semiconductor eutectic systems
(Au-Si, Au-Ge, Al-Ge and Ag-Si). According to this report, for the Au-Si
system, uniformly mixed amorphous layer with composition of
Au$_{71}$Si$_{29}$ was formed. These studies were on continuous thin
films. For nanostructural islands (Au islands on Si) mixing can occur
where ion range is much larger compared to the island thickness
\cite{Satpati2,Satpati3}.\\

\noindent Most of the above mentioned works were done where target films
were continuous and projectile was monomer or cluster ion. We have done
the measurements on nanodispersed target with monomer ion and the catcher
grid being placed very far from target ($\approx$1 cm) at RT. This reduces
the probability of agglomeration of nanoclusters coming out of the target
and deposited on the catcher grid. We observe higher sputtering and higher
probability of crater formation in nanodispersed gold target compare to
thick semi-continuous and continuous target due to MeV self-ion
irradiation and propose it as an experimental evidence of the energy spike
confinement effect \cite{Satpati1,Satpati4}. An energy spike within the
nanoislands which can result either in a thermal spike or a shock wave
could see a spatial confinement effects in nanoislands. Kissel and
Urbassek \cite{Kissel} studied theoretically the effect of 100 keV Au atom
bombardment on spherical Au clusters (radius of 4 nm) using
molecular-dynamics simulation and showed that this bombardment may result
in total disintegration of the clusters.

\section{Experimental}

\noindent In this study, we use single crystal n-type Si(100) substrates
(with native oxide of $\sim$ 2 nm). These substrates were cleaned
sequentially in ultrasonic bath with acetone, methanol, trichloroethylene,
methanol, deionized water and acetone. Au films of thickness 1.3 and 2.6
nm were deposited at a rate of 0.1 nm/s by thermal evaporation under
high-vacuum conditions ($\approx 2\times10^{-6}$ mbar) onto Si at room
temperature (RT).  Ion sputtering experiments were performed mainly with
1.5 MeV Au$^{2+}$ ions with fluence of $1\times10^{14}$ ions cm$^{-2}$ at
different angles of incidence with respect to the surface normal.
Supplementary studies done with 32 keV Au$^{-}$ ions, and 100 MeV
Au$^{8+}$ ions. The ion current for the irradiations was $\sim$ 30 nA for
32 keV and 1.5 MeV irradiation and $\sim$ 15 nA for 100 MeV irradiation
(under secondary electron suppressed geometry). To achieve uniform
irradiation the ion beam was raster over an area of 1$\times$1 cm$^{-2}$.  
During irradiation, $2\times10^{-7}$ mbar pressure was maintained in the
irradiation chamber and sputtered particles were collected on a carbon
coated TEM grid (catcher grid) that was positioned $\approx$ 1 cm above
the target and the catcher surface making an angle ($\approx15^{\circ}$)
with respect to the sample surface.  Transmission electron microscopy
(TEM) measurements were carried out with 200 keV electrons (JEOL
JEM-2010). The cross-section and the planar samples were prepared using
mechanical thinning followed by 3.0 keV Ar ion milling. The amount of
deposited material (the effective film thickness) was measured by
Rutherford backscattering spectrometry (RBS) using 2 MeV He$^{++}$ ions.

\section{Results and Discussions}

Gold initially grows as islands rather than uniform films on the
native-oxide-covered Si substrates. The amount of deposited material is
expressed in terms of an effective thickness, which would be the actual
film thickness if the film were deposited as a film of uniform thickness.
The effective thickness has been determined by using the bulk atomic
density of Au with RUMP simulation package \cite{Doolittle}. The error on
effective thickness determination was $\le$ 10\%. The maximum height of
the Au island is $\sim$30 nm. The particle size and the surface coverage
have been determined from the TEM micrographs with the help of ImageJ
software \cite{ij}. The amount of gold lost from the sample due to
sputtering during irradiation has been calculated from RBS measurements.
Projected range for Au ions in Si and Au for different ion energies is
shown in Table 1.\\

\begin{table}[htbp]
\caption{Projected range for Au ions in Si and Au for different ion 
energies (using the SRIM 2003 range calculation \cite{srim}).}
\begin{center}
\begin{tabular}{|c|c|c|}
\hline
Au ion energy & Projected range in Si & Projected range in Au \\
\hline
32 keV & 23.4 nm & 5.6 nm \\
\hline
1.5 MeV & 360 nm & 100 nm\\
\hline
100 MeV & 14.7 $\mu$m & 5.2 $\mu$m\\
\hline
\end{tabular}
\end{center}
\end{table}

\noindent Au films of thickness 1.3 nm and 2.6 nm was used for the present
study. Fig. 1(a) shows a plan-view TEM image for as-deposited 2.6 nm Au
films on native-oxide/Si substrate with 44\% surface coverage of islands.
Figs. 1(b), (c) and (d) show the TEM images of the sputtered particles
collected from samples like the one shown in Fig. 1(a), which were
irradiated with 32 keV, 1.5 MeV and 100 MeV Au ions, respectively, at a
fluence of $1\times10^{14}$ ions cm$^{-2}$ at normal incidence. From these
figures, it is evident that in case of low energy (keV) ions, sputtering
is less compare to high energy (MeV) in this case (sputtering from
nanoisland films). Fig. 1(e) shows a TEM image (from the same region as in
Fig. 1(d)) from the catcher grid obtained by some underfocousing of the
objective lens. The form of the image of the Au particles shows the
spherical nature of the sputtered particles. Fig. 1(f) shows high
resolution image of a sputtered particle obtained from Fig. 1(d). High
resolution image shows crystalline nature of sputtered particles.
Sputtered particles obtained for irradiation with 32 keV and 1.5 MeV are
also crystalline and spherical in nature (data not shown).\\

\noindent The size distribution of the sputtered particles collected on
catcher grids are shown in Fig. 2. The histograms shown in Figs. 2(a), (b)
and (c) have been determined from many TEM images like Figs. 1(b), (c) and
(d) respectively. Fig. 1(a) shows a bimodal size distribution for the
sputtered particles with $\approx$2\% surface coverage of islands for 32
keV ion irradiation. Fig. 1(b) shows monomodal size distribution for the
sputtered particles with $\approx$14\% surface coverage of islands for 1.5
MeV ion irradiation; here the mean size of islands is 7.5 $\pm$ 0.1 nm and
standard deviation in size distribution is 3.7 $\pm$ 0.2 nm. Fig. 1(c)
shows monomodal size distribution for the sputtered particles with
$\approx$9\% surface coverage of islands for 100 MeV ion irradiation; here
the mean island size is 9.6 $\pm$ 0.1 nm and standard deviation in size
distribution is 4.6 $\pm$ 0.2 nm . Though surface coverage of sputtered
particles for 1.5 MeV ion irradiation is more (14\%) than 100 MeV Au ion
irradiation (9\%) but average sputtered particle size is more for 100 MeV
irradiation than 1.5 MeV irradiation.\\

\noindent Figs. 3(a), (b) and (c) show RBS spectra obtained from
as-deposited 2.6 nm Au/Si and same samples irradiated with 32 keV, 1.5 MeV
and 100 MeV Au ions respectively with fluence of $1\times10^{14}$ ions
cm$^{-2}$ at normal incidence. Loss of Au from the sample due to
sputtering can be estimated from these spectrum's. RBS measurements
showing more sputtering in case of 100 MeV Au irradiation by reducing the
total Au signal and less sputtering in case of 32 keV Au irradiation
compared to 1.5 MeV Au ion irradiation.  From the RBS analyses we observe:  
(a) in case of 32 keV irradiation, the effective film thickness reduces
from 2.6 nm to 1.5 nm upon ion irradiation (an effective reduction of 1.1
nm due to irradiation, i.e. an 42\% thickness reduction (relative to the
initial thickness), (b) in case of 1.5 MeV irradiation the effective film
thickness reduces from 2.6 nm to 0.3 nm (an effective reduction of 2.3 nm
occurred due to irradiation, i.e. a 88\% thickness reduction), (c) in case
of 100 MeV irradiation the effective film thickness reduces from 2.6 nm to
0 nm (an effective reduction of 2.6 nm occurred due to irradiation, i.e. a
100\% thickness reduction). Obviously the sputtering yield is higher for
100 MeV ion irradiation. We have calculated relative sputtering yield (Y)
for these ion energies from RBS measurements,\\

Y (at 32 keV) : Y (at 1.5 MeV) : Y (at 100 MeV) = 1 :  2.1 :  2.4\\

\noindent Figs. 4(a), (b) and (c) show XTEM images of 2.6 nm
Au/native-oxide /Si, irradiated with fluence of $1\times10^{14}$ ions
cm$^{-2}$ at 0$^{\circ}$-impact angle with 32 keV Au$^{-}$ ions, 1.5 MeV
Au$^{2+}$ ions and 100 MeV Au$^{8+}$ ions respectively. In case of 32 keV
embedding is less compared to 1.5 MeV case. For 100 MeV, there is no
embedding. We have seen very less sputtering in case of thick continuous
film compare to island thin film \cite{Satpati1}.\\

\noindent Fig. 5(a) show the XTEM images of 1.3 nm Au/native-oxide/Si.
Figs. 5(b), (c) and (d) show the XTEM images of such samples shown in Fig.
5(a) irradiated with 1.5 MeV Au ions at a fluence of $1\times10^{14}$ ions
cm$^{-2}$ at 0$^{\circ}$, 30$^{\circ}$ and 60$^{\circ}$-impact angle,
respectively.  For this thickness there is no embedding for 0$^{\circ}$
-impact and full embedding for 60$^{\circ}$-impact.  Some embedding can be
seen for 30$^{\circ}$-impact angle as well.  From these results, we may
infer that embedding is more for higher impact angles. However, we have
not checked whether it is more or not if impact angle is higher than
60$^{\circ}$. From the contrast near the silicon surface seen in Fig.
4(a), 4(b), 5(c) and 5(d) it appears that some Au has been pushed into
silicon. In order to demonstrate that the contrast is indeed due to
incorporation of Au into silicon, we have done aqua-regia treatment and
carried out RBS measurements, which proves beyond doubt that indeed Au has
entered into silicon \cite{Satpati3}.\\

\noindent To understand more about the mixing and phase formation, we have
done high resolution lattice imaging from the mixed region (data not
shown). The lattice image shows a d spacing of 0.224 $\pm$ 0.01 nm for
pristine Au/Si (with native oxide) system, which is closer to (111)
interplanar spacing of pristine and bulk Au ions. Upon irradiation with 32
keV and 1.5 MeV Au ions at a fluence of $1\times10^{14}$ ions cm$^{-2}$
and at an impact angle of 0$^{\circ}$, 30$^{\circ}$ and 60$^{\circ}$, the
lattice spacing is found to be 0.293 $\pm$ 0.01 nm.  This value does not
match with any of the interplanar spacing available for the pure gold.
Thus, we conclude that this must belong to a metastable phase of Au/Si
system. The mixed phase reported here is crystalline in nature
\cite{Satpati3}.\\

\noindent From Figs. 4(b) and 5(b) it is evident that embedding of Au into
Si at 0$^{\circ}$ largely depends on Au film thickness. From our earlier
measurement \cite{Satpati3} we have seen that for 0$^{\circ}$ as well as
for 60$^{\circ}$-impact there is no embedding in case of semi-continuous
(94\% film coverage) and in thick continuous Au film on Si for irradiation
at a fluence of $1\times10^{14}$ ions cm$^{-2}$.  In all the cases
whenever there is a embedding (Fig. 4(a), 4(b), 5(c), 5(d)), embedded Au
form a reacted material gold-silicide in the sub surface region.

\section {Conclusions}

We studied keV-MeV Au ion-irradiation effects on isolated nanoislands
which were grown on Si substrates. We have collected sputtered particle on
catcher grid. From TEM measurement sputtering is more in MeV energy region
than keV region. RBS results confirm the effect through enhanced reduction
in thickness for the film irradiated with 1.5 MeV and 100 MeV Au ions in
comparison with that of 32 keV irradiation. We have provided an
experimental evidence for an ion beam induced material push-in (or
burrowing) for 32 keV and 1.5 MeV Au ion irradiation in nanoislands and
nanosilicide formation in nano-Au/Si and its absence in case of 100 MeV
irradiation. The formation of nanosilicide may be useful in fabricating
embedded nanostructures such as nanocontacts and Schottky barriers. The
embedding of nanoislands inside Si has been studied as a function of
impact angle. Embedding is prominent at higher impact angles than at
normal incidence.

\newpage

Figure Captions:

\begin{figure}[htbp]
\caption{\label{Fig1}Plan-view TEM images: (a) as-deposited 2.6 nm thick
Au film on native oxide coated silicon substrate; (b), (c) and (d)
correspond to sputtered particles collected from samples (a), which were
irradiated with 32 keV, 1.5 MeV and 100 MeV Au ions, respectively with a
fluence of $1\times10^{14}$ ions cm$^{-2}$ at normal incidence
(0$^{\circ}$-impact angle); (e) shows sputtered particles collected from
the same region as in (d)) by some underfocousing of the objective lens;
(f) shows high resolution image of a sputtered particle obtained after 100
MeV Au ion irradiation.} \end{figure}

\begin{figure}[htbp]
\caption{\label{Fig2}Size distribution of the sputtered particle collected
on catcher grid following (a) 32 keV, (b) 1.5 MeV, and (c) 100 MeV Au ion
irradiation with ion fluence of $1\times10^{14}$ ions cm$^{-2}$ at 
normal incidence.} \end{figure}

\begin{figure}[htbp]
\caption{\label{Fig3}RBS spectra obtained from as-deposited and
ion-irradiated samples (at a fluence of $1\times10^{14}$ ions
cm$^{-2}$ at normal incidence) for (a) 32 keV, (b) 1.5 MeV, and 
(c) 100 MeV Au ion irradiation.} \end{figure}

\begin{figure}[htbp]
\caption{\label{Fig4}XTEM images of ion irradiated 2.6 nm
Au/native-oxide/Si with fluence of $1\times10^{14}$ ions cm$^{-2}$ at
normal incidence: (a) 32 keV Au$^{-}$ ions, (b) 1.5 MeV Au$^{2+}$ ions and
(c) 100 MeV Au$^{8+}$ ions.} \end{figure}

\begin{figure}[htbp]
\caption{\label{Fig5}XTEM images of 1.3 nm Au film on native-oxide/Si
substrate following 1.5 MeV Au$^{2+}$ bombardment with ion fluence of
$1\times10^{14}$ ions cm$^{-2}$: (a) as-deposited, (b) 0$^{\circ}$-impact
angle, (c) 30$^{\circ}$-impact angle, (d) 60$^{\circ}$-impact angle.}
\end{figure}


\newpage

\begin{thebibliography}{99}

\bibitem{Sigmund} P. Sigmund, in Inelastic Ion-Surface Collisions, edited
by N. H. Tolk, J. C. Tully, W. Heiland, and C. W. White (Academic Press,
New York, 1977), p. 128.

\bibitem{Bay}H. L. Bay, H. H. Anderson, W. O. Hofer, and O. Nielsen, Nucl.
Instrum.  Methods Phys. Res. 132, 301 (1976); P. Sigmund, Appl. Phys.
Lett. 25, 169 (1974);  ibid. 27, 521 (1975).

\bibitem{Ming}L. Yu Ming, in R. Behrisch, and K. Wittmaack (Eds.),
Sputtering by Particle Bombardment III, Springer, Berlin- Heidelberg,
1991, p. 91, and references cited therein.

\bibitem{Lau1}W. M. Lau, Nucl. Instr. Meth. B 16, 41 (1986).

\bibitem{Hofer} W. O. Hofer and H. Gnaser, Nucl. Instr. Meth. B 18, 605
(1987).

\bibitem{Leyen} D. van Leyen, B. Hagenhoff, E. Niehuis, and A.
Benninghoven, J. Vac. Sci.  Tech. A 7, 1790 (1989).

\bibitem{Blumenthal} R. Blumenthal, K. P. Caffey, E. Furman, B. J.
Garrison, and N. Winograd, Phys. Rev. B 44, 12830 (1991).

\bibitem{Karolewski} M. A. Karolewski and R. G. Cavell, Surf. Sci. 480, 47
(2001).

\bibitem{Belykh} S. F. Belykh, I. A. Wojciechowski, V. V. Palitsin, A. V.
Zinoviev, A.  Adriaens, and F. Adams, Surf. Sci. 488, 141 (2001).

\bibitem{Ziegler} J. F. Ziegler, J. P. Biersack, and U. Littmark, The
Stopping and Range of Ions in Solids, Pergamon Press, New York, 1985.

\bibitem{Birtcher} R. C. Birtcher, S. E. Donnelly and S. Schlutig, Phys.
Rev. Lett. 85 (2000) 4968.

\bibitem{Rehn} L. E. Rehn, R. C. Birtcher, S. E. Donnelly, P. M. Baldo and
L. Funk, Phys. Rev. Lett. 87 (2000) 207601.

\bibitem{Bitensky} I. S. Bitensky and E. S. Parilis, Nucl. Instr. Meth. B
21, 26 (1987).


\bibitem{Andersen} H. H. Andersen, A. Brunelle, S. Della-Negra, J. Depauw,
D. Jacquet, and Y. Le Beyec, J. Chaumont and H. Bernas, Phys. Rev. Lett.
80 (1998) 5433.

\bibitem{Bouneau} S. Bouneau, A. Brunelle, S. Della-Negra, J. Depauw, D.
Jacquet, and Y. Le Beyec, M. Putrart, M. Fallavier, J. C. Poizat and H. H.
Andersen, Phys. Rev. B 65 (2002) 144106.

\bibitem{Kissel} Rolf Kissel and Herbert M. Urbassek, Nucl. Instrum. Meth.
Phys. Res. B 180 (2001) 293.

\bibitem{Baranov} I. Baranov, A. Brunelle, S. Della-Negra, D. Jacquet, S.
Kirillov, Y. Le Beyee, A. Novikov, V. Obnorskii, A. Pchelintsev, K. Wien
and S. Yarmiychuk, Nucl. Instrum. Meth. Phys. Res. B 193 (2002) 809.

\bibitem{Satpati1} B. Satpati, D. K. Goswami, S. Roy, T. Som, B. N. Dev
and P. V. Satyam, Nucl. Instr. Meth. Phys. Res. B 212 (2003) 332.

\bibitem{Shapiro} M. H. Shapiro and T. A. Tombrello, Nucl. Instrum. Meth.
Phys. Res. B 152 (1999) 221.

\bibitem{Tsaur} B. Y. Tsaur, S. S. Lau, and J. W. Mayer, Appl. Phys. Lett. 36
(1979) 168.

\bibitem{Poate} J. M. Poate and T. C. Tisone, Appl. Phys. Lett. 24 (1974)
391.

\bibitem{Lau} S. S. Lau, B. Y. Tsaur, M. von Allmen, J. W. Mayer, B.
Stritzker, C. W. White, and B. Appleton, Nucl. Instr. Meth. Phys. Res.
182/183 (1981) 97.

\bibitem{Satpati2} B. Satpati, P. V. Satyam, T. Som, and B. N. Dev, J. Appl.
Phys. 96 (2004) 5212.

\bibitem{Satpati3} B. Satpati, P. V. Satyam, T. Som, and B. N. Dev, Appl.
Phys. A 79 (2004) 447.

 \bibitem{Satpati4}B. Satpati, D. K. Goswami, U. D. Vaishnav, T. Som, B.
N. Dev and P. V. Satyam, Nucl. Instr. Meth. Phys. Res. B 212 (2003) 157.

\bibitem{Doolittle} L. R. Doolittle, Nucl. Instrum. Meth. Phys. Res. B 9
(1985) 344.

\bibitem{ij} http://rsb.info.nih.gov/ij/

\bibitem{srim} http://www.srim.org/

\end{thebibliography}
\end{document}